\documentclass[aps,twocolumn,preprintnumbers,amsmath,amssymb,superscriptaddress,showpacs,longbibliography]{revtex4-2}

\usepackage[colorlinks,bookmarks=true,citecolor=blue,linkcolor=red,urlcolor=blue,citecolor=blue]{hyperref}

\usepackage{graphicx}
\usepackage{color}
\usepackage{lineno}
\usepackage{tikz}
\tikzset{every picture/.style={line width=0.75pt}}
\usepackage{enumitem}
\usepackage{adjustbox}
\usepackage{pifont}

\def\figref#1{Fig.~\ref{#1}}

\begin{document}

\switchlinenumbers
\title{Thermodynamics of the Hubbard Model on the Bethe Lattice}

\author{Jia-Lin Chen}
\altaffiliation{These authors contributed equally to this work.}
\affiliation{Beijing National Laboratory for Condensed Matter Physics and Institute of Physics,
Chinese Academy of Sciences, Beijing 100190, China.}
\affiliation{School of Physical Sciences, University of Chinese Academy of Sciences, Beijing 100049, China.}

\author{Zhen Fan}
\altaffiliation{These authors contributed equally to this work.}
\affiliation{Beijing National Laboratory for Condensed Matter Physics and Institute of Physics,
Chinese Academy of Sciences, Beijing 100190, China.}
\affiliation{School of Physical Sciences, University of Chinese Academy of Sciences, Beijing 100049, China.}

\author{Bo Zhan}
\affiliation{Beijing National Laboratory for Condensed Matter Physics and Institute of Physics,
Chinese Academy of Sciences, Beijing 100190, China.}
\affiliation{School of Physical Sciences, University of Chinese Academy of Sciences, Beijing 100049, China.}

\author{Jiahang Hu}
\affiliation{Beijing National Laboratory for Condensed Matter Physics and Institute of Physics,
Chinese Academy of Sciences, Beijing 100190, China.}
\affiliation{School of Physical Sciences, University of Chinese Academy of Sciences, Beijing 100049, China.}

\author{Tong Liu}
\affiliation{Beijing National Laboratory for Condensed Matter Physics and Institute of Physics,
Chinese Academy of Sciences, Beijing 100190, China.}
\affiliation{School of Physical Sciences, University of Chinese Academy of Sciences, Beijing 100049, China.}

\author{Junyi Ji}
\affiliation{Beijing National Laboratory for Condensed Matter Physics and Institute of Physics,
Chinese Academy of Sciences, Beijing 100190, China.}
\affiliation{School of Physical Sciences, University of Chinese Academy of Sciences, Beijing 100049, China.}

\author{Kang Wang}
\affiliation{Beijing National Laboratory for Condensed Matter Physics and Institute of Physics,
Chinese Academy of Sciences, Beijing 100190, China.}
\affiliation{School of Physical Sciences, University of Chinese Academy of Sciences, Beijing 100049, China.}

\author{Hai-Jun Liao}
\affiliation{Beijing National Laboratory for Condensed Matter Physics and Institute of Physics,
Chinese Academy of Sciences, Beijing 100190, China.}
\affiliation{Songshan Lake Materials Laboratory, Dongguan, Guangdong 523808, China.}

\author{Tao Xiang}
\email{txiang@iphy.ac.cn}
\affiliation{Beijing National Laboratory for Condensed Matter Physics and Institute of Physics, Chinese Academy of Sciences, Beijing 100190, China.}
\affiliation{School of Physical Sciences, University of Chinese Academy of Sciences, Beijing 100049, China.}
\affiliation{Beijing Academy of Quantum Information Sciences, Beijing, China.}

\begin{abstract}
 We investigate the thermodynamic properties of the Hubbard model on the Bethe lattice with a coordination number of 3 using the thermal canonical tree tensor network method. Our findings reveal two distinct thermodynamic phases: a low-temperature antiferromagnetic phase, where spin SU(2) symmetry is broken, and a high-temperature paramagnetic phase. A key feature of the system is the separation of energy scales for charge and spin excitations, which is reflected in the temperature dependence of thermodynamic quantities and the disparity between spin and charge gaps extracted from their respective susceptibilities. At the critical point, both spin and charge susceptibilities exhibit singularities, suggesting that charge excitations are not fully decoupled from their spin counterparts. Additionally, the double occupancy number exhibits a non-monotonic temperature dependence, indicative of an entropy-driven Pomeranchuk effect. These results demonstrate that the loopless Bethe lattice effectively captures the essential physics of the Hubbard model while providing a computationally efficient framework for studying strongly correlated electronic systems.
\end{abstract}

\maketitle


\section{INTRODUCTION} \label{sec:Intro}

 The Hubbard model \cite{hubbardElectronCorrelationsNarrow1997} is one of the fundamental models in condensed matter physics, providing profound insights into the intricate interplay between spin and charge degrees of freedom. It has been instrumental in explaining a variety of emergent phenomena, including antiferromagnetic order, Mott metal-insulator transition \cite{imadaMetalinsulatorTransitions1998, mottMetalInsulatorTransitions2004}, charge-spin separation \cite{haldaneLuttingerLiquidTheory1981}, and stripe phases \cite{whiteStripes6LegHubbard2003,hagerStripeFormationDoped2005,mizusakiGaplessQuantumSpin2006,zhengStripeOrderUnderdoped2017,qinAbsenceSuperconductivityPure2020}. As a result, the Hubbard model serves as a fundamental framework for exploring the physics of strongly correlated electrons. In particular, the doped two-dimensional Hubbard model has attracted significant attention due to its close connection to the high-temperature superconductivity observed in copper oxides \cite{bednorzPossibleHighTcSuperconductivity1986, orensteinAdvancesPhysicsHighTemperature2000}.

  Despite its foundational importance, a comprehensive understanding of the Hubbard model has primarily been confined to one-dimensional \cite{liebAbsenceMottTransition1968} and infinite-dimensional \cite{metznerCorrelatedLatticeFermions1989,muller-hartmannCorrelatedFermionsLattice1989,georgesPhysicalPropertiesHalffilled1993} systems, where exact solutions are available. Considerable progress has also been made in intermediate dimensions through numerical simulations \cite{whiteStripes6LegHubbard2003,hagerStripeFormationDoped2005,mizusakiGaplessQuantumSpin2006,zhengStripeOrderUnderdoped2017,qinAbsenceSuperconductivityPure2020,hirschTwodimensionalHubbardModel1985,duffySpecificHeatTwodimensional1997,leblancSolutionsTwoDimensionalHubbard2015,schaferTrackingFootprintsSpin2021b,ivTwodimensionalHubbardModel2022,liTangentSpaceApproach2023,staudtPhaseDiagramThreedimensional2000,kozikNeelTemperatureThermodynamics2013,iskakovPhaseTransitionsPartial2022,lenihanEvaluatingSecondOrderPhase2022,songMagneticThermodynamicDynamical2025,sinhaFinitetemperatureTensorNetwork2022,tangFinitetemperaturePropertiesStrongly2013,khatamiEffectParticleStatistics2012,khatamiThreedimensionalHubbardModel2016}, yet accurately determining low-temperature properties in the thermodynamic limit remains challenging \cite{sinhaFinitetemperatureTensorNetwork2022,tangFinitetemperaturePropertiesStrongly2013,khatamiEffectParticleStatistics2012,khatamiThreedimensionalHubbardModel2016}. Quantum Monte Carlo methods suffer from the notorious sign problem \cite{lohSignProblemNumerical1990, troyerComputationalComplexityFundamental2005,wuSufficientConditionAbsence2005,liSolvingFermionSign2015,weiMajoranaPositivityFermion2016, liSignProblemFreeFermionicQuantum2019,alexandruComplexPathsSign2022} away from half-filling. Density-matrix renormalization \cite{whiteDensityMatrixFormulation1992c} or, more generally, tensor network methods \cite{xiangDensityMatrixTensor2023} are constrained by the area law of entanglement entropy, limiting their application to large or low-temperature systems. Dynamical mean-field theory (DMFT) \cite{georgesDynamicalMeanfieldTheory1996a} captures local correlations. It becomes rigorous in the infinite-dimension limit but significantly truncates spatial correlations, which are crucial for understanding the physics of the low-dimensional Hubbard model. 

 The Hubbard model has also been studied on the Bethe lattice, focusing on its ground-state properties. Early work \cite{lepetitDensitymatrixRenormalizationStudy2000} applied the density matrix renormalization group (DMRG) to the Cayley tree, a finite-size structure that shares the local connectivity of the Bethe lattice. These studies found that the antiferromagnetic (AF) order increases monotonically with the Hubbard interaction at half-filling. Recently, the uniform fermion tree tensor network method has been employed to investigate the Hubbard model directly on the Bethe lattice, both at and away from half-filling \cite{luntsHubbardModelBethe2021}. The findings reveal an antiferromagnetic insulating phase at half-filling, which transitions into a paramagnetic metallic phase upon doping, with a first-order metal-insulator transition separating the two. Furthermore, the metallic phase exhibits Landau Fermi liquid behavior, characterized by a nonzero quasiparticle weight. Despite these advances, the thermodynamic properties of the Hubbard model on the Bethe lattice remain unexplored.

 The Bethe lattice is a loop-free tree graph \cite{katsuraBetheLatticeBethe1974}. Its unique tree-like structure reduces the impact of non-local correlations and provides an ideal setup for simplifying complex many-body problems, validating the Bethe approximation \cite{betheStatisticalTheorySuperlattices1997, katsuraBetheLatticeBethe1974} and other theoretical approaches, and enabling efficient numerical treatments. These properties have played a crucial role in studying critical phenomena in models such as the Ising \cite{katsuraBetheLatticeBethe1974}, Potts \cite{bingCorrelationFunctionPotts2000}, and Bose-Hubbard models \cite{semerjianExactSolutionBoseHubbard2009b}, spin glasses \cite{mezardBetheLatticeSpin2001, thoulessSpinGlassBetheLattice1986, kopecShortrangeInteractionIsing2006, laumannCavityMethodQuantum2008}, and localization effects \cite{abou-chacraSelfconsistentTheoryLocalization1973, mirlinLocalizationTransitionAnderson1991, altshulerQuasiparticleLifetimeFinite1997, savitzAndersonLocalizationBethe2019}, avoiding the exponential computational costs typically associated with such problems. Moreover, the Bethe lattice, particularly in the infinite-dimensional limit, allows exact solutions to impurity problems within the DMFT framework \cite{georgesDynamicalMeanfieldTheory1996a}. The absence of loops and the resulting simplification of Green's functions and self-energies further enhance its utility in studying quantum phase transitions, critical exponents, and scaling behaviors.

 The loopless structure of the Bethe lattice also makes it an ideal framework for tensor network techniques, as local tensors can be readily and reliably canonicalized  \cite{liEfficientSimulationInfinite2012, quThermalTensorNetwork2019}. This property facilitates the fruitful application of tree tensor network renormalization group methods to various quantum many-body problems \cite{nagajQuantumTransversefieldIsing2008, nagySimulatingQuantumSystems2012, liEfficientSimulationInfinite2012, quThermalTensorNetwork2019}. For instance, in the XXZ model \cite{quThermalTensorNetwork2019}, tree tensor network calculations reveal not only a first-order phase transition from the XY phase to the (anti)ferromagnetic phase at the ground state, but also a finite-temperature transition from a paramagnetic phase to a U(1) symmetry-breaking phase, without generating gapless Goldstone modes \cite{laumannAbsenceGoldstoneBosons2009, quThermalTensorNetwork2019}. Furthermore, the correlation length of the XXZ model, and similarly for other models, remains finite even at the critical point \cite{liEfficientSimulationInfinite2012,quThermalTensorNetwork2019,semerjianExactSolutionBoseHubbard2009b,nagySimulatingQuantumSystems2012,xiangDensityMatrixTensor2023}. Specifically, in the XXZ model, the correlation length is bounded by $\xi = 1/\ln(z-1)$, where $z$ is the coordination number of the Bethe lattice \cite{liEfficientSimulationInfinite2012,quThermalTensorNetwork2019}.

 In this work, we extend the thermal canonical tree tensor network method \cite{quThermalTensorNetwork2019} to fermionic systems and apply it to the Hubbard model on the $z=3$ Bethe lattice. Our findings reveal different thermodynamic behaviors attributed to spin-charge separation. Moreover, by leveraging the canonical form of the tree tensor network ansatz, we observe clear signatures of spin-charge separation and spontaneous SU(2) symmetry breaking in the entanglement spectrum. 
 
 The paper is organized as follows. In Sec.~\ref{sec:ModelAndMethod}, we provide a brief introduction to the Hubbard model and the thermal canonical tree tensor network method. Section \ref{sec:NumericalResults} presents the numerical results for the thermodynamics of the half-filled Hubbard model on the Bethe-lattice. Finally, in Sec.~\ref{sec:SummaryAndDiscussion}, we summarize our findings and discuss potential research directions.


\begin{figure*}[t]
    \centering
    \includegraphics[width=\linewidth]{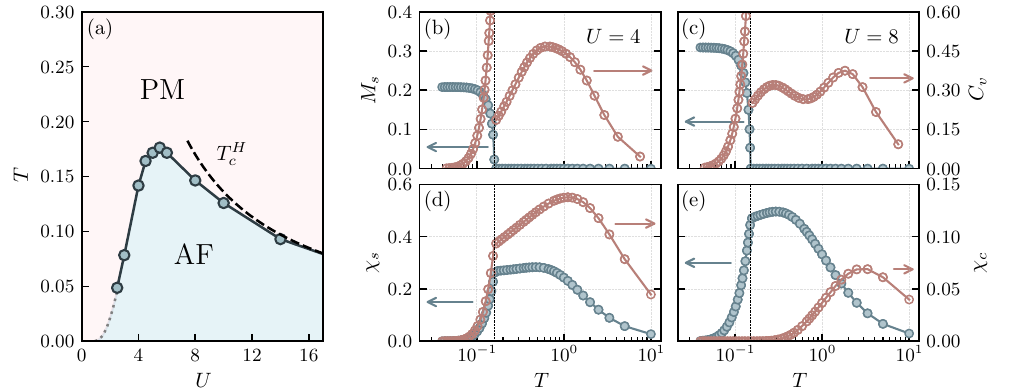}
    \caption{ (a) Phase diagram of the Hubbard model at half-filling. The dashed line represents the critical temperature of the Heisenberg model, $T_c^H = 1.36/U$, while the dotted line shows an exponential extrapolation to the small $U$ limit. Temperature dependence of staggered magnetization $M_s$ and specific heat $C_v$ (b-c), along with spin susceptibility $\chi_s$ and charge susceptibility $\chi_c$ (d-e), for the Hubbard model at $U = 4$ (middle panels) and $U = 8$ (right panels), respectively. The dotted lines indicate the antiferromagnetic phase transition temperatures $T_c$. 
    \label{fig_PhaseDigram}}
\end{figure*}

\section{MODEL AND METHOD} \label{sec:ModelAndMethod}

The Hubbard model is described by the Hamiltonian:
\begin{equation}
    \label{eq:HubbardModel}
    H = -t \sum_{\langle ij \rangle \sigma} c^{\dagger}_{i\sigma} c_{j\sigma} + U \sum_{i} n_{i\uparrow} n_{i\downarrow} - \mu \sum_{i\sigma} n_{i\sigma},
\end{equation}
where $c_{i\sigma}^\dagger$ ($c_{i\sigma}$) creates (annihilates) an electron with spin $\sigma$ at site $i$, and $n_{i\sigma} = c_{i\sigma}^\dagger c_{i\sigma}$ is the corresponding number operator. The coefficient $t$ is the hopping integral, $U$ is the on-site Hubbard interaction, and $\mu$ is the chemical potential. For convenience, we set $t=1$ as the energy unit. At half-filling, the chemical potential is given by $\mu = U / 2$, as dictated by particle-hole symmetry. 

 Thermodynamic properties of the Hubbard model are encoded in its partition function 
 \begin{equation}
     Z = \mathrm{Tr} \rho (\beta ) , \qquad \rho(\beta ) = e^{-\beta H}, 
 \end{equation} 
 where $\rho$ is the density operator and $\beta = 1/ T$ is the inverse of temperature $T$. 

 In our calculation, we represent $\rho$ as a tree tensor network operator defined on the $z=3$ Bethe lattice, starting from the infinite temperature limit $\beta =0$. At this limit, each local tensor is initialized as an equal superposition of the operators corresponding to the four possible states at a lattice site, specifically, $(|0\rangle \langle0| + |\uparrow\rangle \langle \uparrow | + |\downarrow\rangle \langle \downarrow | + |\uparrow\downarrow \rangle \langle \uparrow\downarrow |) / 2$. We perform the imaginary evolution of the density matrix to determine its value at a finite $\beta$ \cite{quThermalTensorNetwork2019}. At each step of evolution, we canonicalize the local tensors to minimize truncation errors, ensuring that the density operator is optimally calculated within the tensor network framework \cite{jiangAccurateDeterminationTensor2008b, liEfficientSimulationInfinite2012}. Fermionic statistics are incorporated by employing swap gates \cite{corbozFermionicMultiscaleEntanglement2009, corbozSimulationStronglyCorrelated2010}. 

 A detailed description of the algorithm, including the canonicalization process and the implementation of swap gates, is provided in Appendix \ref{appendix:A}. Our code implementation is publicly available (see Ref. \cite{sourceCodeThermalBetheLattice}). Benchmark results for the exactly solvable model of non-interacting fermions ($U = 0$) are presented in Appendix \ref{appendix:B}.


\section{Results}\label{sec:NumericalResults}

  The results presented below are obtained using the thermal tree tensor network method \cite{quThermalTensorNetwork2019} with a bond dimension of $\chi = 200$, unless stated otherwise. To balance accuracy and computational efficiency, we employ both the first-order ($\tau_1 = 0.05$) and fourth-order ($\tau_4 = 0.1$) Suzuki-Trotter decompositions \cite{suzukiSTTransformationApproachAnalytic1987a, inoueSTTransformationApproachAnalytic1988, ostmeyerOptimisedTrotterDecompositions2023}.

\subsection{Thermal phase diagram }

 The ground state of the Hubbard model on the $z=3$ Bether Lattice is known to exhibit antiferromagnetic order \cite{luntsHubbardModelBethe2021}. Our thermodynamic calculation reveal that this antiferromagnetic order emerges at a finite temperature when $U$ is finite. Thus, the Hubbard model undergoes a finite temperature transition from the high-temperature paramagnetic phase to the low-temperature antiferromagnetic phase. Figure~\ref{fig_PhaseDigram}(a) shows the thermal phase diagram of the Hubbard model, while Figure~\ref{fig_PhaseDigram}(b-e) illustrates the singularity of the spin susceptibility and other thermodynamic quantities used to determine the Néel temperature $T_c$. 
 
 The antiferromagnetic transition appears to be continuous within the limits of numerical accuracy. Below $T_c$, the spin SU(2) symmetry is spontaneously broken. 
 However, the Goldstone mode, which is typically associated with low-lying excited states resulting from continuous symmetry breaking, does not emerge in this system. This absence is due to the fact that the Goldstone theorem does not hold on the Bethe lattice \cite{laumannAbsenceGoldstoneBosons2009, quThermalTensorNetwork2019}. The breaking of the SU(2) symmetry indicates that the phase transition is predominantly driven by spin fluctuations.

 The Néel temperature marks the phase boundary between the paramagnetic and antiferromagnetic phases. It exhibits distinct behaviors in the small- and large-U limits.  In the small-$U$ regime, the Néel temperature increases rapidly with $U$. From previous calculations \cite{luntsHubbardModelBethe2021}, it is known that the ground state is in an antiferromagnetically ordered insulating phase where the staggered magnetization follows an exponential dependence, $M_s \sim \exp(a U)$ ($a$ is a $U$-independent coefficient), for small $U$. Consequently, $T_c$ is expected to be exponentially small but increase exponentially with $U$ in this limit. Accurately resolving this exponentially small $T_c$ requires significantly enlarging the bond dimension of local tensors, which is computationally prohibitive due to high CPU and memory costs. Instead, by extrapolating the values of $T_c$ obtained for moderately small $U$ using the function $T_c \sim \exp(- \alpha / U)$, we estimate the exponent to be $\alpha \sim 7.7$, as shown by the dotted curve shown in Figure~\ref{fig_PhaseDigram}(a). 
 
 In the large-$U$ regime, the low-energy spin excitations can be effectively described by the Heisenberg model, with an exchange coupling $J_{\text{eff}} = 4t^2 / U$. The critical temperature predicted by the Heisenberg model is  $T_c^H = 1.36 / U$ \cite{quThermalTensorNetwork2019}, which we include in Figure~\ref{fig_PhaseDigram}(a) for comparison. Our numerical results agree well with the Heisenberg prediction, confirming the expected behavior in this regime.

\begin{figure}[b]
    \centering
    \includegraphics[]{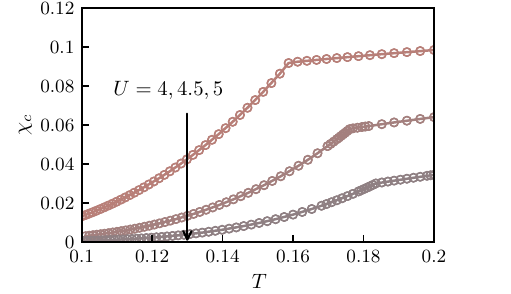}
    \caption{Charge susceptibility $\chi_c$ as a function of temperature $T$ for the Hubbard model on the $z = 3$ Bethe-lattice with $U = 4, 4.5$, and $5$. \label{fig_ChargeSusSingularity}}
\end{figure}

 In addition to the antiferromagnetic phase transitions evident in the thermodynamic quantities shown in Fig. \ref{fig_PhaseDigram}(b-e) at $T_c$, the specific heat $C_v$ exhibits distinct behavior in the high-temperature regime preceding the transition. Specifically, the number of peaks in the specific heat varies with the interaction strength $U$. For the $U=8$ case, $C_v$ exhibits a double-peak structure, whereas for $U=4$, only a single peak is observed. This double-peak structure is a manifestation of spin-charge separation, a phenomenon widely observed in the Hubbard model \cite{chandraNearlyUniversalCrossing1999, glockeHalffilledOnedimensionalExtended2007, duffySpecificHeatTwodimensional1997, tangFinitetemperaturePropertiesStrongly2013, khatamiThreedimensionalHubbardModel2016}. This separation is further evident when comparing the temperature dependence of spin susceptibility $\chi_s$ with that of charge susceptibility $\chi_c$ for the $U=8$ case.

 Notably, the charge susceptibility also displays singular behavior at $T_c$, which is particularly pronounced in the $U=4$ case, as shown in Fig. \ref{fig_PhaseDigram}(d). The concurrent emergence of singularities in both spin and charge channels suggests that the spin and charge collective excitations remain coupled, even though clear signatures of spin-charge separation are observed above $T_c$. In contrast to the spin susceptibility, the singularity in the charge susceptibility gradually diminishes with increasing $U$, as evidenced in Fig. \ref{fig_ChargeSusSingularity}.  For $U=8$, the large Mott insulating gap strongly suppresses the charge susceptibility, making it difficult to identify the charge singularity in the data presented in Fig. \ref{fig_PhaseDigram}(e). Resolving this singularity in $\chi_c$ at $U=8$ would require a bond dimension $\chi$ much higher than 200, which is technically demanding.

\begin{figure}[t]
    \centering
    \includegraphics[]{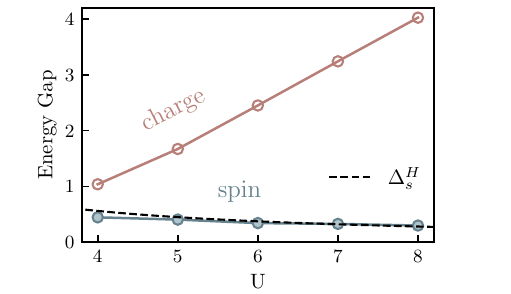}
    \caption{Spin and charge excitation gaps, $\Delta_s$ and $\Delta_c$, as functions of $U$. The spin and charge gaps are obtained by fitting the exponential decay behaviors of the spin and charge susceptibilities at low temperatures with the formula $\chi_s \sim \exp(-\Delta_s/T)$ and $\chi_c \sim \exp(-\Delta_c/2T)$. The dashed line $\Delta_s^H$ represents the spin excitation gap of the Heisenberg model with an effective coupling constant $J = 4t^2/U$.  \label{fig_ChargeGap}}
\end{figure}

  Figure \ref{fig_ChargeGap} shows the spin gap $\Delta_s$ and charge gap $\Delta_c$ as functions of $U$. These gaps are extracted from the exponential decay of the spin and charge susceptibilities at low temperatures. Over the range of interactions considered in this work,  $\Delta_s$ decreases monotonically with increasing $U$, converging toward the Heisenberg limit $\Delta_s \sim 4t^2 \Delta_s^H/U$, where $\Delta_s^H \sim 0.56$ is the spin gap of the Heisenberg model \cite{quThermalTensorNetwork2019}. In contrast, $\Delta_c$ increases linearly with $U$, consistent with the results obtained from ground-state calculations \cite{luntsHubbardModelBethe2021}. This distinct behavior further highlights the contrasting roles of spin and charge excitations. 

 \begin{figure}[t]
     \centering
     \includegraphics[]{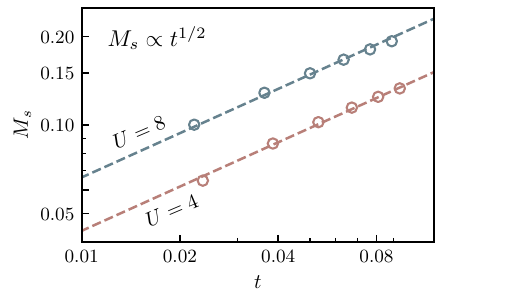}
     \caption{Scaling behavior of the staggered magnetization $M_s$ as a function of the reduced temperature $t = \left(T_c - T\right)/T_c$ for the Hubbard model with $U=4$ and $U=8$. The dashed lines represent the fitting curves obtained using the formula $ M_s = C t^{1/2} $ with $C$ a fitting parameter.\label{fig_CriticalExponent}}
 \end{figure}

 To characterize the critical behavior near the antiferromagnetic transition, we conduct a scaling analysis of the order parameter. Figure \ref{fig_CriticalExponent} shows the temperature dependence of the staggered magnetization $M_s$ just below $T_c$ for the Hubbard model with $U=4$ and $U=8$. In  both cases, the numerical data for $M_s$ are well described by the scaling law:
 \begin{equation}
     M_s \sim \left( \frac{T_c - T}{T_c} \right)^\alpha ,
 \end{equation}
 where the critical exponent $\alpha \approx 1/2$. This value of the critical exponent aligns with the predictions of mean-field theory \cite{quThermalTensorNetwork2019,semerjianExactSolutionBoseHubbard2009b}, reflecting the unique geometric properties of the Bethe lattice. The Bethe lattice, with its loopless and tree-like structure, effectively mimics an infinite-dimensional system where fluctuations are suppressed.
 
\subsection{{Pomeranchuk effect}} 

\begin{figure}[t]
    \centering
    \includegraphics[width=\linewidth]{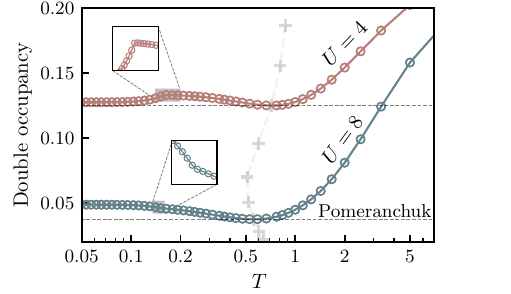}
    \caption{Double occupancy $D$ as a function of temperature for the Hubbard model at $U = 4$ and $8$. The minima of $D$ indicate the onset temperature of the Pomeranchuk effect. Cross markers connected by the dashed line show how these minima vary with $U$ with a $\Delta U = 1$ step. Insets provide enlarged views of $D$ near the critical temperatures. 
    \label{fig_MIT_and_Pomeranchuk_Effect}}
\end{figure}

 An interesting feature observed at intermediate-high temperatures is the non-monotonic behavior of the double occupancy number $D \equiv \langle n_{i\uparrow} n_{i\downarrow} \rangle $. Specifically, as shown in \figref{fig_MIT_and_Pomeranchuk_Effect}, $D$ first decreases and then increases, forming a minimum at a temperature denoted as $T_m$. This reentrant phenomenon is a manifestation of the so-called Pomeranchuk effect \cite{wernerInteractionInducedAdiabaticCooling2005}. The Pomeranchuk effect, originally observed in liquid $^3$He, is a phenomenon where entropy drives an unexpected phase transition \cite{richardsonPomeranchukEffect1997}. In the context of the Hubbard model, it signifies a crossover from an entropy-driven localized phase to an energetically favored itinerant phase. This effect has been observed in DMFT and other numerical simulations across various lattice geometries and dimensions \cite{georgesPhysicalPropertiesHalffilled1993, liTangentSpaceApproach2023, khatamiEffectParticleStatistics2012, khatamiThreedimensionalHubbardModel2016}. 

 The Pomeranchuk effect arises as an entropy-driven phenomenon induced by strong correlations in the Hubbard model. Coulomb repulsion suppresses double occupancy in the strong coupling limit, leading to a state dominated by local magnetic moments. These moments contribute significantly to the spin entropy at intermediate temperatures, favoring localization. However, as the temperature drops below the characteristic exchange interaction $J\sim 4t^2/U$, spin fluctuations and kinetic energy become more prominent, favoring itinerancy and increasing double occupancy.

\begin{figure}[b]
    \centering
    \includegraphics[]{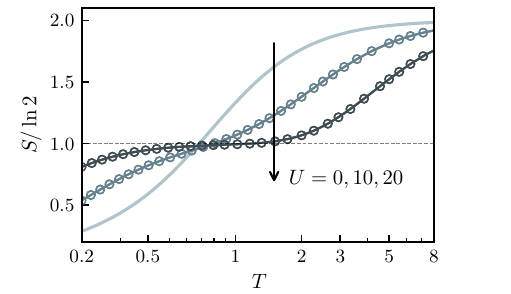}
    \caption{Thermal entropy $S$ vs $T$ for the Hubbard model with different $U$. The bond dimension $\chi=80$. \label{fig_Entropy}}
\end{figure}

\begin{figure*}[t]
    \centering
    \includegraphics[width=\linewidth]{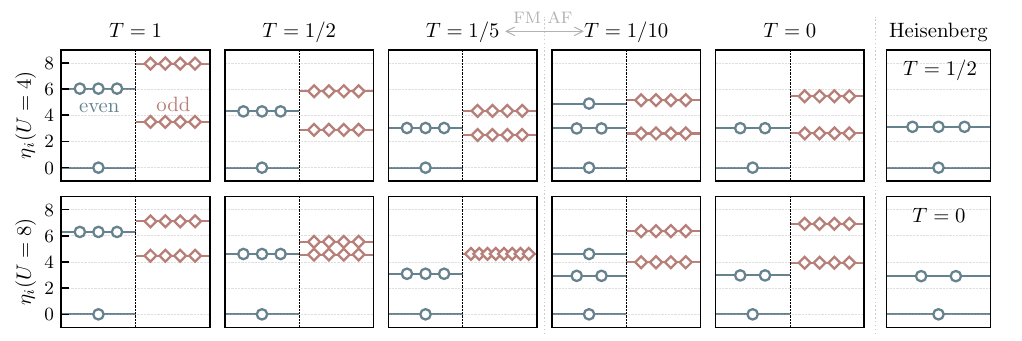}
    \caption{Entanglement spectrum of $\rho(\beta / 2)$ at various temperatures for $U=4$ (upper panels) and $U=8$ (lower panels). The rightmost panel presents reference results from the Heisenberg model. The plots depict the two lowest spectral levels in the even-parity (circles) and odd-parity (diamonds) sectors.
    \label{fig_Entanglement_Spectrum}}
\end{figure*}

 This effect can also be understood through the Maxwell relation between the thermal entropy $S$ and the double occupancy
\begin{equation} 
    \left( \frac{\partial D}{\partial T} \right)_U = -\left( \frac{\partial S}{\partial U} \right)_T .
\end{equation} 
 As shown in Fig.~\ref{fig_Entropy}, the thermal entropy exhibits two distinct behaviors in the limits of small and large $U$. In the small $U$ limit, the system behaves as a Fermi liquid, and the entropy increases almost linearly with temperature near the minimum in  $S$. In contrast, in the large $U$ limit, the system becomes insulating in the same temperature range, and the entropy develops a plateau as a consequence of charge-spin separation, approaching a constant value $\ln d$, where $d=2$ is the dimension of a localized spin. Consequently, the derivative of entropy with respect to $U$ shows distinct behavior: at low temperatures, $\partial S /\partial U > 0$, while at high temperatures, $\partial S /\partial U  < 0$. As a result, the double occupancy exhibits non-monotonic temperature dependence, initially suppressed at high temperatures due to strong correlations, but increasing at lower temperatures as the system gains energy through delocalization.

In \figref{fig_MIT_and_Pomeranchuk_Effect}, the evolution of ($T_m$, $D_m$) with $U$ is also depicted. As $U$ increases, $T_m$ initially declines, reaching a minimum around $U \sim 6$, before rising again. This trend in $T_m$ is more consistent with the linked cluster expansion results \cite{khatamiEffectParticleStatistics2012, khatamiThreedimensionalHubbardModel2016} than with the DMFT prediction \cite{georgesPhysicalPropertiesHalffilled1993}, which suggests a monotonic decrease in $T_m$ with $U$, becoming vanishingly small for $U \geq 4$.

 The double occupancy $D$ exhibits a kink at the antiferromagnetic transition temperature $T_N$ (see the insets in \figref{fig_MIT_and_Pomeranchuk_Effect}). This slope discontinuity across the antiferromagnetic critical point arises from the interplay between local correlations and long-range spin fluctuations. Once long-range antiferromagnetic order sets in, it imposes tighter constraints on the available spin configurations, affecting the charge distribution and $D$. Below $T_N$, $D$ steadily decreases with decreasing temperature for $U=4$, whereas it increases at lower temperatures for $U=8$. Similar contrasting behavior has been reported in cellular DMFT calculations \cite{fratinoSignaturesMottTransition2017a}.

\subsection{Entanglement Spectrum}

  To further understand the nature of charge-spin separation across the critical temperature, we have calculated the entanglement spectrum of the density operator defined by 
\begin{equation}
    \eta_i = -2\ln \lambda_i , 
\end{equation}
 where $\lambda_i$ is the singular spectrum of the thermal density matrix $\rho (\beta /2)$, and it can be directly extracted from the bond matrix. The spectrum is normalized such that the largest singular value $\lambda_0 = 1$, ensuring $\eta_0 = 0$. In our calculations, fermion number parity conservation is enforced on each local tensor. This enables us to distinguish and separately analyze the entanglement spectrum in the even- and odd-parity sectors in accordance with the imposed $Z_2$ symmetry. The parity-odd levels $\eta_i^{\text{odd}}$ originate from charge fluctuations or fermion hopping between neighboring sites, which alter the fermion number parity on the bond connecting the two sites. In contrast, the parity-even levels $\eta_i^{\text{even}}$ arise from spin fluctuations across the bond, which preserve the fermion number parity but exhibit spin SU(2) symmetry.
  
 \figref{fig_Entanglement_Spectrum} presents the entanglement spectra $\eta_i$ across a range of temperatures, spanning the critical point. Throughout the entire temperature range, from high to low temperatures, the lowest spectral level $\eta_1^{\text{odd}}$ in the odd parity sector remains four-fold degenerate within numerical precision. This observation indicates that the symmetry associated with charge fluctuations remains unbroken, aligning with the absence of any transition driven by charge fluctuations across the whole temperature range. In contrast, the lowest spectral level $\eta_1^{\text{even}}$ in the even parity sector exhibits three-fold degeneracy at high temperatures, reflecting the presence of spin SU(2) symmetry. However, this symmetry is broken below the antiferromagnetic transition temperature, causing the three-fold degenerate level to split into a two-fold degenerate level and a non-degenerate level. These split levels correspond to the two transverse lower-energy excited modes and one longitudinal excited mode, respectively.

 At high temperatures, the lowest level in the odd-parity sector, $\eta_1^{\mathrm{odd}}$, remains lower than its counterpart in the even-parity sector,  $\eta_1^{\text{even}}$. However, in the strong coupling limit and at low temperatures, this ordering reverses, with $\eta_1^{\mathrm{odd}}$ rising above $\eta_1^{\mathrm{odd}}$. The temperature at which this inversion occurs decreases with decreasing $U$, consistent with the behavior observed in thermodynamic quantities at intermediate temperatures. At these temperatures, spin fluctuations remain strong, while charge fluctuations are significantly suppressed due to the opening of the Mott insulating gap. For small $U$, we find that $\eta_1^{\mathrm{odd}}$ is always lower than $\eta_1^{\mathrm{even}}$. The critical $U$ below which this ordering persists is found to be $U \sim 4.96$. 

\section{SUMMARY AND DISCUSSION}\label{sec:SummaryAndDiscussion}

 In summary, we have investigated the thermodynamic properties of the half-filled Hubbard model on a Bethe lattice with coordination number $z=3$, utilizing the fermionic tree tensor network renormalization group method. Our numerical results reveal two distinct phases: a high-temperature paramagnetic phase and a low-temperature antiferromagnetic ordered phase. The critical temperature separating these phases increases with $U$ in the weak coupling regime, peaks around $U \sim 5.5$, and declines as $U$ continues to increase.

 The interplay between kinetic hopping and on-site Coulomb interaction leads to a distinct separation in the characteristic energy scales of collective charge and spin excitations. This separation is clearly reflected in thermodynamic observables. For instance, the specific heat exhibits two well-separated peaks at large $U$. Moreover, charge and spin susceptibilities exhibit markedly different behavior. In the antiferromagnetic phase, the spontaneous breaking of spin SU(2) symmetry is confirmed through both local magnetization and entanglement spectrum analysis. However, magnon excitations in this ordered phase remain gapped, highlighting the unique geometric properties of the Bethe lattice, where long-range magnetic order persists without the Goldstone modes typically present in regular bipartite lattices.

 In the intermediate-temperature regime, the Pomeranchuk effect manifests universally across all $U$, as evidenced by the non-monotonic evolution of the double occupancy number with temperature. This effect arises from the competition between entropy-driven localized spins and energy-driven itinerant electron correlations, further highlighting the intricate interplay between charge and spin excitations.

 Our results exhibit striking qualitative agreement with the studies of the Hubbard model in three-dimensional and higher-dimensional lattices \cite{staudtPhaseDiagramThreedimensional2000, kozikNeelTemperatureThermodynamics2013,iskakovPhaseTransitionsPartial2022, lenihanEvaluatingSecondOrderPhase2022, khatamiThreedimensionalHubbardModel2016,songMagneticThermodynamicDynamical2025,georgesPhysicalPropertiesHalffilled1993,shaoAntiferromagneticPhaseTransition2024}, demonstrating that the Bethe lattice faithfully captures the essential physics of the Hubbard model beyond one dimension. The loop-free structure of the Bethe lattice, combined with the fermionic tree tensor network algorithm, enables computationally efficient and numerically precise simulations of strongly correlated systems, circumventing the exponential complexity inherent to conventional methods.

 A promising direction for future research is to extend our work beyond half-filling in the Hubbard model and explore other interacting fermion models, such as the $t-J$ model \cite{zhangEffectiveHamiltonianSuperconducting1988} and the $t-U-J$ model \cite{xiangIntrinsicElectronHole2009}, to investigate the emergence of novel phenomena on the Bethe lattice. Given that the Bethe lattice provides an easy access framework to the calculation of entanglement spectra, it is intriguing to examine how these spectra characterize such phenomena. Moreover, extending the tree tensor network method to study excited states \cite{ostlundThermodynamicLimitDensity1995a, haegemanVariationalMatrixProduct2012a, vanderstraetenExcitationsTangentSpace2015b} represents another important avenue, as excitation spectra offer valuable insights into the system's low-energy physics \cite{chiSpinExcitationSpectra2022a, chiDynamicalSpectraSpin2024, liuResolvingGeometricExcitations2024, wangFractionalizationSignaturesDynamics2024, zhuEmergentMajoranaMetal2024, chenSpinExcitationsShastrySutherland2024}.


\section*{ACKNOWLEDGMENTS}

 This work is supported by the Innovation Program for Quantum Science and Technology (Grant No.~2021ZD0301800), the National Natural Science Foundation of China (Grants No.~12488201, No.~12322403 and No.~12347107), the Strategic Priority Research Program of Chinese Academy of Sciences (Grants No.~XDB0500202), the National Key Research and Development Project of China (Grants No.~2024YFA1408604 and 2022YFA1403900) and the Youth Innovation Promotion Association of Chinese Academy of Sciences (Grant No.~2021004).

\clearpage
\appendix

\renewcommand{\theequation}{A.\arabic{equation}}
\setcounter{equation}{0}

\section{Thermal tree tensor network renormalization}\label{appendix:A} 

\subsection{Fermionic tree tensor network ansatz}

\begin{figure}[t]
    \centering
    \input{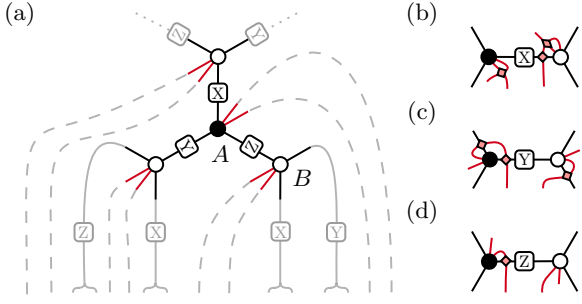}
    \caption{Tree tensor network ansatz for the $z=3$ Bethe lattice and the corresponding swap gates required for updating different bonds. (a) The ansatz; (b)-(d) Swap gates for updating the X, Y, and Z bonds, respectively. \label{fig_lattice_and_ansatz}}
\end{figure}

 In this work, we represent the density operator $\rho=\exp(-\beta H)$ as a tree tensor network (TTN) and determine it by taking the imaginary time evolution. By employing the first-order Trotter-Suzuki decomposition formula (or higher-order variants), the density operator can be approximated as
\begin{eqnarray}
   e^{-\tau H} & \approx & e^{-\tau H_x} e^{-\tau H_y} e^{-\tau H_z} + \mathcal{O}(\tau^2), \\
   H_\alpha & = & \sum_i h_{i, i+\alpha} , \qquad (\alpha = \hat{x}, \hat{y}, \hat{z} ), 
\end{eqnarray}
 where $\tau$ is a small Trotter parameter, and $H_\alpha$ represent the Hamiltonian component along the $\alpha$-direction. Since all terms within $H_\alpha$ commute with each other, $[h_{i,i+\alpha}, h_{i^\prime, i^\prime+\alpha}]=0$, the exponential of $H_\alpha$ can be factorized as
\begin{equation}
   e^{-\tau H_\alpha} = \prod_{i} e^{-\tau h_{i,i+\alpha}}.
\end{equation}
 In practical simulations, $\tau$ serves as the discrete imaginary time step used to iteratively perform the imaginary time evolution. This process begins at infinite temperature, where the density operator reduces to the identity operator, and proceeds by incrementally evolving the system to finite temperatures. This approach provides a systematic and numerically tractable framework for constructing the density operator within the TTN formalism.

\begin{figure}[t]
    \centering
    \input{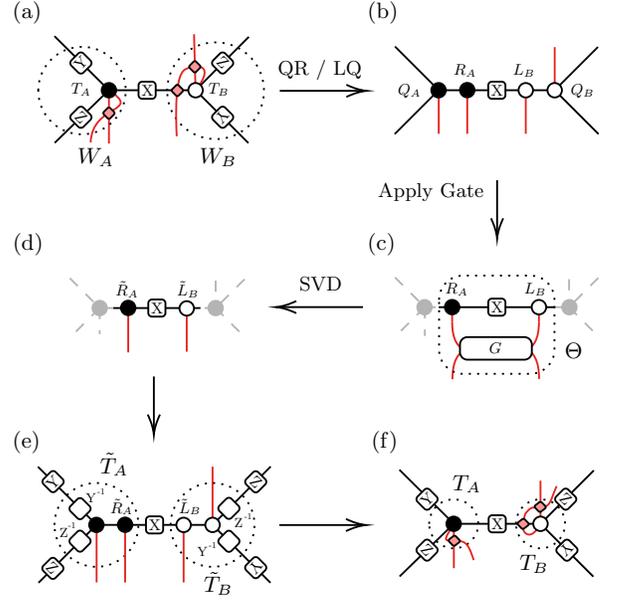}
    \caption{Algorithm flow\label{fig_update_scheme}}
\end{figure}

 To account for the sign change arising from the exchange of two fermion operators in the Hubbard model, we introduce the swap gate, defined as
\begin{equation}
    \begin{array}{l}
    \begin{tikzpicture}[x=0.75pt,y=0.75pt,yscale=-1,xscale=1]
    \draw [line width=0.75]    (209.94,160.35) -- (260.29,130.24) ;
    \draw    (210.06,130) -- (260.53,160.59) ;
    \draw  [color={rgb, 255:red, 0; green, 0; blue, 0 }  ,draw opacity=1 ][fill={rgb, 255:red, 255; green, 150; blue, 150 }  ,fill opacity=1 ][line width=0.75]  (228.87,141.73) -- (241.36,141.73) -- (241.36,148.86) -- (228.87,148.86) -- cycle ;
    
    \draw (198.2,120.46) node [anchor=north west][inner sep=0.75pt]  [font=\normalsize]  {$i$};
    \draw (264.82,120.46) node [anchor=north west][inner sep=0.75pt]  [font=\normalsize]  {$j$};
    \draw (198.2,155.07) node [anchor=north west][inner sep=0.75pt]  [font=\normalsize]  {$k$};
    \draw (264.82,155.07) node [anchor=north west][inner sep=0.75pt]  [font=\normalsize]  {$l$};
\end{tikzpicture}
    \end{array} \equiv \  \delta_{i,l}\  \delta_{j,k}\  S(i,j), \label{eq:swap_gate}
\end{equation}
where $S(i,j)$ is a sign factor:
\begin{equation}
    S(i,j) \equiv \left\{
    \begin{array}{ll}
    -1 & \text{if\ \ } p(i) = p(j) = -1 \\
    +1 & \text{if\ \ otherwise.}
    \end{array}
    \right. ,
\end{equation}
and $p(i) = \pm 1$ represents the fermion parity quantum number associated with the physical or virtual basis state $i$. The swap gate ensures the correct anti-symmetry under fermion exchange, which is essential for preserving the fermionic statistics in the simulation.

 As an example, let us consider the update procedure for an $x$-bond evolution. We use $\lambda_\alpha$ ($\alpha = x,y,z)$) to represent the bond vector along the $\alpha$-direction, which effectively describes the entanglement spectra across the corresponding bond. The workflow is illustrated in \figref{fig_update_scheme}, with the detailed steps outlined as follows:
\begin{enumerate}[label=(\alph*)]
    \item Absorb the bond tensors $\lambda_y$ and $\lambda_z$ into the site tensors $T_i\ (i=A, B)$ to construct the $W_i$ tensor. Then, permute the indices to ensure the physical indices (where the gate will act) are adjacent, introducing swap gates as required to account for fermionic sign changes.
    
    \item Perform a QR decomposition $W_A = Q_A R_A$, where $Q_A$ is a unitary matrix and $R_A$ an upper triangular matrix. Similarly, perform an LQ decomposition $W_B = L_B Q_B$, where $Q_B$ is a unitary matrix and $L_B$ a lower triangular matrix. These decompositions effectively reduce the computational cost.
    
    \item Construct the $\Theta$ tensor by applying the imaginary time evolution gate $G = \exp(-\tau H)$ to the adjacent physical indices.
    
    \item Perform a singular value decomposition (SVD) on $\Theta$, truncating the virtual bond basis space to retain the largest $\chi$ singular values. The truncated singular value vector is then used to update the bond tensor $\lambda_x$. 
    
    \item Update the site tensor $T_A$ by multiplying $\tilde{R}_A$ (obtained by truncating the $x$-bond dimension of $R_A$ to $\chi$) with $Q_A$, and split off the bond tensors $\lambda_y$ and $\lambda_z$ by multiplying their inverses. The other site tensor $T_B$ is updated similarly. 
    
    \item Restore the physical indices to their original positions by permuting them back, introducing swap gates as necessary to account for fermionic sign changes.
\end{enumerate}
 This step-by-step procedure ensures an efficient and accurate update while maintaining the correct structure of the tensor network.

\subsection{Canonicalization}
 To minimize the truncation errors in the imaginary-time evolution, the TTN should be canonized prior to applying the local evolution operators. This canonization process involves solving the following self-consistent equations \cite{liEfficientSimulationInfinite2012, tindallGaugingTensorNetworks2023}: 
\begin{equation}
    \begin{array}{l}
    
\begin{adjustbox}{width=0.7\linewidth}
\begin{tikzpicture}[x=0.75pt,y=0.75pt,yscale=-1,xscale=1]
    \draw    (151.5,170.75) .. controls (152.57,200.21) and (150.75,201.25) .. (165.75,207.25) ;
    \draw    (151.5,170.75) .. controls (152.57,141.29) and (150.75,140.25) .. (165.75,134.25) ;
    \draw  [fill={rgb, 255:red, 255; green, 255; blue, 255 }  ,fill opacity=1 ] (135.3,170.75) .. controls (135.3,179.7) and (142.55,186.95) .. (151.5,186.95) .. controls (160.45,186.95) and (167.7,179.7) .. (167.7,170.75) .. controls (167.7,161.8) and (160.45,154.55) .. (151.5,154.55) .. controls (142.55,154.55) and (135.3,161.8) .. (135.3,170.75) -- cycle ;

    \draw    (97.2,170.7) .. controls (96.25,280.15) and (144.25,295.15) .. (169.75,281.65) ;
    \draw    (97.2,170.7) .. controls (96.25,61.25) and (144.25,46.25) .. (169.75,59.75) ;
    \draw  [fill={rgb, 255:red, 255; green, 255; blue, 255 }  ,fill opacity=1 ] (81,170.8) .. controls (81,179.75) and (88.25,187) .. (97.2,187) .. controls (106.15,187) and (113.4,179.75) .. (113.4,170.8) .. controls (113.4,161.85) and (106.15,154.6) .. (97.2,154.6) .. controls (88.25,154.6) and (81,161.85) .. (81,170.8) -- cycle ;
    
    \draw    (186.5,69.98) -- (246.65,112.28) ;
    \draw [line width=0.75]    (300.25,112.28) -- (246.65,112.28) ;
    \draw  [fill={rgb, 255:red, 255; green, 255; blue, 255 }  ,fill opacity=1 ] (206.68,91.13) .. controls (206.68,85.38) and (211.33,80.73) .. (217.08,80.73) .. controls (222.82,80.73) and (227.48,85.38) .. (227.48,91.13) .. controls (227.48,96.87) and (222.82,101.53) .. (217.08,101.53) .. controls (211.33,101.53) and (206.68,96.87) .. (206.68,91.13) -- cycle ;
    \draw    (246.65,112.28) -- (180,130.47) ;
    \draw  [fill={rgb, 255:red, 255; green, 255; blue, 255 }  ,fill opacity=1 ] (195.68,122.63) .. controls (195.68,116.88) and (200.33,112.23) .. (206.08,112.23) .. controls (211.82,112.23) and (216.48,116.88) .. (216.48,122.63) .. controls (216.48,128.37) and (211.82,133.03) .. (206.08,133.03) .. controls (200.33,133.03) and (195.68,128.37) .. (195.68,122.63) -- cycle ;
    \draw    (246.65,229.23) .. controls (220.25,199.75) and (219.25,140.25) .. (246.65,112.28) ;
    \draw    (186.5,271.53) -- (246.65,229.23) ;
    \draw [line width=0.75]    (299.25,229.23) -- (246.65,229.23) ;
    \draw  [fill={rgb, 255:red, 255; green, 255; blue, 255 }  ,fill opacity=1 ] (206.67,250.38) .. controls (206.67,256.12) and (211.33,260.78) .. (217.08,260.78) .. controls (222.82,260.78) and (227.48,256.12) .. (227.48,250.38) .. controls (227.48,244.63) and (222.82,239.98) .. (217.08,239.98) .. controls (211.33,239.98) and (206.67,244.63) .. (206.67,250.38) -- cycle ;
    \draw    (246.65,229.23) -- (180,211.03) ;
    \draw  [fill={rgb, 255:red, 255; green, 255; blue, 255 }  ,fill opacity=1 ] (195.67,218.88) .. controls (195.67,224.62) and (200.33,229.28) .. (206.08,229.28) .. controls (211.82,229.28) and (216.48,224.62) .. (216.48,218.88) .. controls (216.48,213.13) and (211.82,208.48) .. (206.08,208.48) .. controls (200.33,208.48) and (195.67,213.13) .. (195.67,218.88) -- cycle ;
    \draw    (415.2,170.8) .. controls (414.6,92.8) and (415.8,85.2) .. (447.8,84.8) ;
    \draw    (415.2,170.7) .. controls (414.6,248.7) and (415.8,256.3) .. (447.8,256.7) ;
    \draw  [fill={rgb, 255:red, 255; green, 255; blue, 255 }  ,fill opacity=1 ] (399,170.7) .. controls (399,179.65) and (406.25,186.9) .. (415.2,186.9) .. controls (424.15,186.9) and (431.4,179.65) .. (431.4,170.7) .. controls (431.4,161.75) and (424.15,154.5) .. (415.2,154.5) .. controls (406.25,154.5) and (399,161.75) .. (399,170.7) -- cycle ;
    
    \draw    (246.65,229.23) .. controls (273.05,199.75) and (274.05,140.25) .. (246.65,112.28) ;
    \draw  [fill={rgb, 255:red, 255; green, 255; blue, 255 }  ,fill opacity=1 ] (238.78,112.28) .. controls (238.78,107.93) and (242.3,104.4) .. (246.65,104.4) .. controls (251,104.4) and (254.53,107.93) .. (254.53,112.28) .. controls (254.53,116.62) and (251,120.15) .. (246.65,120.15) .. controls (242.3,120.15) and (238.78,116.62) .. (238.78,112.28) -- cycle ;
    \draw  [fill={rgb, 255:red, 255; green, 255; blue, 255 }  ,fill opacity=1 ] (238.77,229.23) .. controls (238.77,224.88) and (242.3,221.35) .. (246.65,221.35) .. controls (251,221.35) and (254.52,224.88) .. (254.52,229.23) .. controls (254.52,233.57) and (251,237.1) .. (246.65,237.1) .. controls (242.3,237.1) and (238.77,233.57) .. (238.77,229.23) -- cycle ;
    \draw  [dash pattern={on 0.84pt off 2.51pt}]  (169.75,59.75) -- (186.5,69.98) ;
    \draw  [dash pattern={on 0.84pt off 2.51pt}]  (169.75,281.65) -- (186.5,271.53) ;
    \draw  [dash pattern={on 0.84pt off 2.51pt}]  (165.75,134.25) -- (180,130.47) ;
    \draw  [dash pattern={on 0.84pt off 2.51pt}]  (165.75,207.25) -- (180,211.03) ;


\draw (320,157) node [anchor=north west][inner sep=0.75pt]  [scale=1/0.6]  {$=\hspace{0.5em} \lambda_\alpha$};

\draw (82,162) node [anchor=north west][inner sep=0.75pt]  [scale=1.25]  {$M^{y}_{\bar\alpha}$};

\draw (137,162) node [anchor=north west][inner sep=0.75pt]  [scale=1.25]  {$M^{z}_{\bar\alpha}$};

\draw (400,162) node [anchor=north west][inner sep=0.75pt]  [scale=1.25]  {$M^{x}_{\alpha}$};

\draw (209.5,84) node [anchor=north west][inner sep=0.75pt]  [scale=1.25]  {$Y$};

\draw (197.5,115.5) node [anchor=north west][inner sep=0.75pt]  [scale=1.25]  {$Z$};

\draw (197, 211.5) node [anchor=north west][inner sep=0.75pt]  [scale=1.25]  {$Z$};

\draw (209, 244) node [anchor=north west][inner sep=0.75pt]  [scale=1.25]  {$Y$};

\draw (250, 244) node [anchor=north west][inner sep=0.75pt]  [scale=1/0.6]  {$T_\alpha ^\dagger$};
\draw (250, 80) node [anchor=north west][inner sep=0.75pt]  [scale=1/0.6]  {$T_\alpha$};

\end{tikzpicture}
\end{adjustbox}
    \end{array}
\end{equation}
where $M^i_\alpha$ ($i=x,y,z$, $\alpha = A, B$) are the message tensors. Index $\bar \alpha$ is complementary to $\alpha$: if $\alpha = A$, then $\bar \alpha = B$, and vice versa. 

 Once the symmetric and semi-positive definite message tensor is obtained, the TTN can be transformed into its canonical form by inserting the gauge transformation $U^i_\alpha$, which is derived from the square root of the message tensor:
\begin{equation} 
    M^i_\alpha = U^i_\alpha U^{i\dagger}_\alpha. 
\end{equation}
 For further technical details, please refer to Ref. \cite{tindallGaugingTensorNetworks2023} and Appendix A.2 of Ref. \cite{quThermalTensorNetwork2019}.

\renewcommand{\theequation}{B.\arabic{equation}} 
\setcounter{equation}{0}

\section{Benchmark results for the U=0 case} \label{appendix:B}
\begin{figure}[t]
    \centering
    \includegraphics[width=\linewidth]{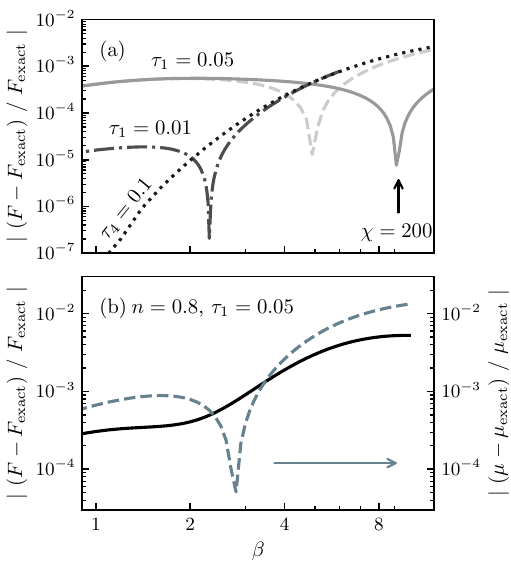}
    \caption{Comparison between TNN results ($\chi=100$) and exact solutions for the non-interacting fermion model ($U=0$). (a) Relative error of the free energy, $ (F - F_{\mathrm{exact}})/F_{\text{exact}}$, at half-filling; (b) Relative error of the free energy and chemical potential, $(\mu - \mu_{\mathrm{exact}})/\mu_{\mathrm{exact}}$, at a filling of $n = 0.8$.
    The results were obtained using the first-order Trotter-Suzuki decomposition scheme, except for the curve labeled $\tau_4$, which was calculated using the fourth-order Trotter-Suzuki decomposition formula.
    \label{fig_benchmark_z=3_free_fermion}}
\end{figure}

 In the $U=0$ limit, the Hubbard model becomes exactly soluble. The Green's function for free fermions on the Bethe lattice, $G$, is explicitly derived in Refs. \cite{mahanEnergyBandsBethe2001, ecksteinHoppingBetheLattice2005}. Using this result \cite{ecksteinHoppingBetheLattice2005}, we obtain the density of states as
\begin{equation}
    \rho(\omega) = -\frac{1}{\pi} \Im G_{11}(\omega +i0) = \frac{1}{2\pi} \frac{\sqrt{4-\omega^2}}{p-\omega^2/z},
\end{equation}
where $p=z/(z-1)$. The free energy is then calculated as
\begin{equation}
    F=\frac{2}{\beta} \int \ln\left( 1 + e^{-\beta \omega} \right)\rho\left(\omega \right)\ d\omega ,
\end{equation}
 where the factor of 2 accounts for spin degeneracy. Despite its simplicity, the free fermion model exhibits high entanglement in real space, making it a challenging yet ideal platform for benchmarking our method. 

 \figref{fig_benchmark_z=3_free_fermion} compares our TTN results for the free energy and other physical quantities with the exact solutions for this non-interacting fermion system. In the high-temperature regime, the primary source of error is the Trotter error, which can be mitigated by reducing the Trotter parameter $\tau$. In the low-temperature regime, the truncation error becomes dominant, and this can be reduced by increasing the bond dimension $\chi$. 
 
 To further clarify the truncation error, we performed a calculation using fourth-order Trotter decomposition with $\tau_4 = 0.1$, where the cumulative Trotter error scales as $\sim \tau^4$ and is negligible within the temperature range of interest. As shown in the figure, the results obtained with the first-order Trotter-Suzuki decomposition approach to the error curve of the fourth-order decomposition after the sharp drop, confirming that truncation error dominates at low temperatures.

 
\bibliography{BetheLatticeHubbardBib}

\end{document}